\documentclass[aps,twocolumn,showpacs]{revtex4}
\usepackage{graphicx}
\usepackage{tabularx}
\usepackage{color}

\begin{document}

\title{Ambiguities of the Rate of Oxygen Formation During Stellar Helium Burning \\ in the $^{12}$C$(\alpha,\gamma)$ Reaction.}

\author{Moshe Gai}
\affiliation{LNS at Avery Point, University of Connecticut, Groton, CT 06340-6097 \\
 and WNSL, Dept of Physics, Yale University, New Haven, CT 06520-8124}
 
\begin{abstract}
The rate of oxygen formation determines the C/O ratio during stellar helium burning. It is the single most important nuclear input of stellar evolution theory including the evolution of Type II and Type Ia supernova. Yet the low energy cross section of the fusion of $^4$He + $^{12}$C denoted as the $^{12}$C$(\alpha,\gamma)^{16}$O reaction still remains uncertain after forty years of extensive work. We analyze and critically review the most recent measurements of complete angular distributions of the outgoing gamma-rays at very low energies ($E_{cm}  \geq  1.0$ MeV). Our analysis of the angular distributions measured with the EUROGAM/GANDI arrays leads to considerably larger error bars than published which excludes them from the current sample of "world data". We show that the current sample of "world data" of the measured $E2$ cross section factors below 1.7 MeV cluster into two distinct groups that lead to two distinct extrapolations of $S_{E2}(300) \ \approx 60$ or $\approx 154$ keVb. We point to a much neglected discrepancy between the measured $E1-E2$ phase difference ($\phi_{12}$) and unitarity as required by the Watson theorem, suggesting systematic problem(s) in some of the measured gamma-ray angular distributions. The ambiguity of the extrapolated $S_{E2}(300)$ together with a previously observed ambiguity of $S_{E1}(300)$ represent the current state of the art of the field. These discrepancies must be resolved by future measurements of complete and detailed angular distributions of the $^{12}$C$(\alpha,\gamma)$ reaction at very low energies ($E_{cm}  \leq  1.0$ MeV).

\end{abstract}

\preprint{UConn-40870-00XX}

\maketitle

Stellar Helium burning that follows Hydrogen burning in stars is an important stage in the evolution of stars. During this stage the elements carbon and oxygen are formed and as such it is one of the most vivid examples of the anthropic principle \cite{Fowler}. During this stage carbon is synthesized by the so called "triple alpha process" but at the same time carbon is also destroyed by fusing with an additional alpha-particle to form $^{16}$O in the $^{12}$C$(\alpha,\gamma)^{16}$O reaction. Hence the formation of oxygen in stellar helium burning determines the C/O ratio; an essential parameter in stellar evolution theory \cite{Fowler}. 

\begin{figure}
 \includegraphics[width=3.5in]{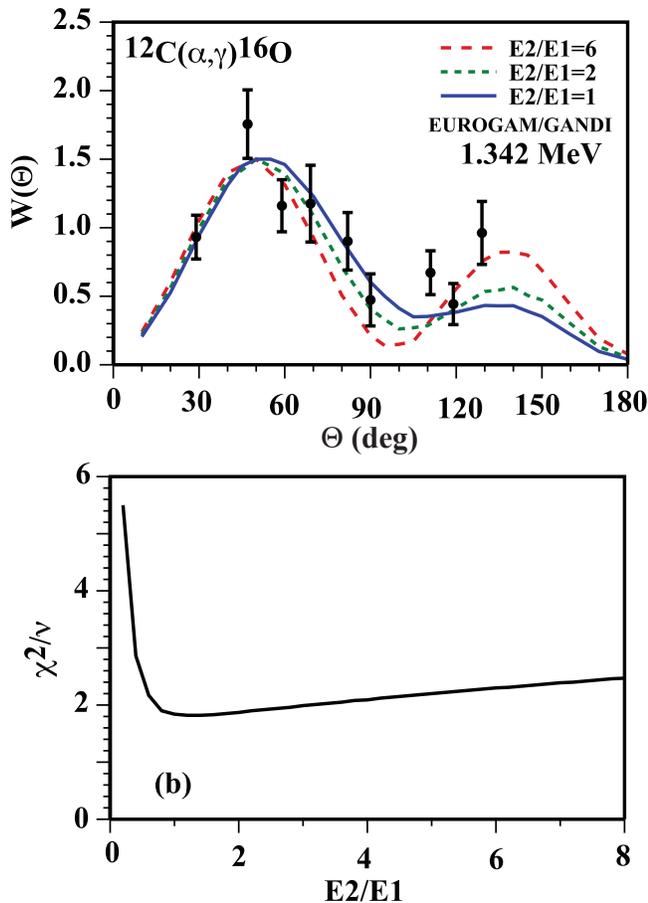}
 \caption{\label{all} (Color Online) (a) The measured angular distribution of the $^{12}$C$(\alpha,\gamma)$ reaction \cite{Ham2} together with the E1 + E2 fits for three values of the E2/E1 ratio as discussed in the text. (b) The reduced $\chi^2/\nu$ obtained for different E2/E1 ratios.}
\end{figure}

The importance of the C/O ratio for the evolution of massive stars ($M  >  8M_\odot$) that evolve to core collapse (type II) supernova has been discussed extensively \cite{Weaver}. More recently it was shown that the C/O ratio is also important for understanding the $^{56}$Ni mass fraction produced by lower mass stars ($M \ \approx \ 1.4M_\odot$) that evolve into Type Ia supernova (SNeIa) \cite{SNeIa}. Thus the C/O ratio is also important for understanding the light curve of SNeIa. Such SNeIa are used as cosmological "standard candles" with which the accelerated expansion of the universe and dark energy were recently discovered \cite{RMP}. 

Stellar evolution theory requires the knowledge of the C/O ratio with an uncertainty of 5\%. This requires accurate measurements at low energies and extrapolation of the measured astrophysical cross section factors to the Gamow window at 300 keV \cite{Fowler}. Since mainly two ($\ell$ = 1 and $\ell$ = 2) partial waves contribute to the reaction, accurate angular distribution data are needed at low energies to determine with high accuracy the astrophysical cross section factors $S_{E1}(300)$ and $S_{E2}(300)$ defined in \cite{Fowler}. This goal has not been achieved as yet. 

Recently some of the most impressive gamma-ray measurements of the $^{12}$C$(\alpha,\gamma)^{16}$O reaction were published \cite{Kunz,Ham1,Ham,Ham2,Plag} including measurements of complete angular distribution at center of mass energies approaching 1.0 MeV. These measurements employ large luminosities of the order of 10$^{35} \ cm^{-2}sec^{-1}$ with integrated luminosities close to one inverse fb. \cite{Ham1,Ham,Ham2,Plag}, and a large (so called $4\pi$) array of gamma-ray detectors (but some of the arrays employ low efficiency HpGe detectors which led in some cases to insufficient counting statistics). Such unprecedented data with unprecedented characteristics led to an expectation of a resolution of the long (now forty years old) debate on the value of the low energy cross section of the $^{12}$C$(\alpha,\gamma)$ reaction. While these data did not resolve the outstanding questions they provide the first possible detailed study of the cross section of the $^{12}$C$(\alpha,\gamma)$ reaction at low energies approaching 1.0 MeV.

In this paper we analyze and critically review these new measurements of angular distributions of gamma-rays from the $^{12}$C$(\alpha,\gamma)$ reaction \cite{Kunz,Ham1,Ham,Ham2,Plag}. We focus our attention on angular distribution data in order to reveal trends in the cross section factors measured at the current lowest energies. Specifically we study the $E2$ cross section factors ($S_{E2}$) measured at energies ($E_{cm}$) below 1.7 MeV in order to avoid the energy region where higher lying ($1^-$ and $2^+$) states dominate and to be most sensitive to the bound $2^+$ state at 6.917 MeV in $^{16}$O that governs the $E2$ cross section at stellar burning energies. We show that the "world data" on $S_{E2}$ below 1.7 MeV cluster into two groups that differ by an average factor of 2.6 and consequently these data extrapolate to two distinct solutions of $S_{E2}(300) \ \approx 60$ or $\approx 154$ keVb. The ambiguity in the value of the extrapolated $S_{E2}(300)$ resembles the previously observed ambiguity in the value of the extrapolated $S_{E1}(300)$ where the small value solution of the E1 cross section factor [$S_{E1}(300) \ \approx \ 10$ keVb] cannot be ruled out \cite{Hale,Gial1}. And we point out a much neglected disagreement of the measured $E1-E2$ relative phase angle ($\phi_{12}$) with unitarity which together with the major disagreement on the value of the "cascade cross section" \cite{Gial2} defines the major challenges facing future measurements in this field. 

We will describe the stringent requirements needed in future studies (see for example \cite{Gai,LUNA-MV}) in order to resolve these ambiguities. The exact values and energy dependence of $S_{E2}$ and $S_{E1}$ are essential for extrapolating the proposed measurements of the total reaction cross section to 300 keV (see for example \cite{Ug13}).

We analyzed all the published angular distributions measured at low energy ($E_{cm}  <  1.5$ MeV) with the GANDI/EUROGAM array at Stuttgart \cite{Ham1,Ham2}. We employed the standard Legendre polynomial expansion as for example shown in Equation 4.3 of \cite{Ham2} and we used the published angular attenuation coefficients. The angular distributions measured at 891 keV and 903 keV, shown in Fig. 4 of \cite{Ham1}, were not included in our analysis since the data point were measured with error bars nearly 100\% (or larger). In order to simplify the analysis we fixed the relative angle ($\phi_{12}$) at the value predicted by equ. (1) discussed below, and we varied only one parameter ($S_{E2}/S_{E1}$) apart from an over all normalization. 

As shown in Fig. 1 the $E2/E1$ ratio at 1.342 MeV can be varied by a factor as large as six and still yield a similar quality fit, with only a slight increase in $\chi ^2/\nu$ from 1.8 to 2.4. The same figure demonstrates that the data points measured at backward angles (larger than $90^\circ$) provide the largest sensitivity to the $E2/E1$ ratio, but these few (three) data points are measured with poor precision considerably worst than 10\%. It is clear from Fig. 1 that precise data (5-10\% statistics) measured with small angular bins (10$^\circ$ or smaller) at large backward angles ($90^\circ - 160^\circ$) are essential for an accurate determination of the $E1$ and $E2$ cross section factors. 

The $\chi ^2$ values shown in Fig. 1(b) yield ${S_{E2} \over S_{E1}}(1.342) \ = \ 1.4^{+1.6}_{-0.6}$ for a fixed value of the relative angle of $\phi_{12} \ = \ 54 ^\circ$ predicted by equ. (1) and discussed below. The $S_{E2}/S_{E1}$ ratios obtained for all other published angular distributions measured at $E_{cm} <  1.5$ MeV \cite{Ham1,Ham2} are shown in Fig. 2.  The large and asymmetric error bars deduced in this analysis are considerably different than those published in \cite{Ham1,Ham,Ham2}. We conclude that the $S_{E2}/S_{E1}$ ratios measured with the EUROGAM/GANDI arrays are not determined with sufficient accuracy (not even 50\%) to define the cross section factors at energies below 1.5 MeV. Thus we do not include these data in the sample of current "world data".

\begin{figure}
 \includegraphics[width=3.5in]{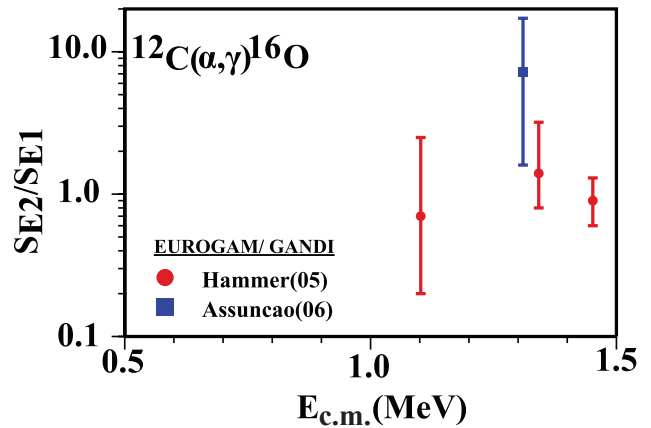}
\caption{\label{20deg} (Color Online) The E2/E1 ratios deduced in the current analysis of the data obtained using the EUROGAM/GANDI arrays \cite{Ham1,Ham2}.}
\end{figure}

\begin{figure}
\includegraphics[width=3.5in]{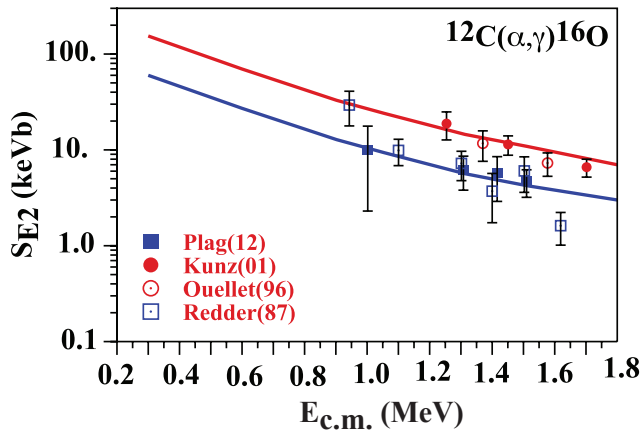}
\caption{\label{no2}(Color Online) The measured $S_{E2}$ values \cite{Kunz,Plag,Redder,Ouellet} and the corresponding R-matrix fits. The two distinct groupings of the data extrapolate to $60 \pm 12$ and $154 \pm 31$ keVb. The $S_{E2}$ values measured using the GANDI \cite{Ham1} and EUROGAM \cite{Ham2} arrays are excluded, as discussed in the text.}
\end{figure}

Excluding the results of the Stuttgart collaboration \cite{Ham1,Ham2} from the sample of world data is in agreement with the finding of Brune and Sayre  \cite{Sayre}, but it is in conflict with Schuermann {\em et al.} \cite{Sch12} that included these data in their sample of world data. In contrast Schuermann {\em et al.} \cite{Sch12} removed the data of Redder {\em et al.} \cite{Redder} and Ouellet {\em et al.} \cite{Ouellet} from their sample of the world data. If so their selection criteria together with the critical review discussed here and in \cite{Sayre} would leave only the recent data of Kunz {\em et al.} \cite{Kunz} and Plag {\em et al.} \cite{Plag} in the current sample of "world data" of measured angular distributions at energies below 1.7 MeV. This is clearly less than a satisfactory situation for such an important cross section.

In Fig. 3 we show the published "world data" of $S_{E2}$ values deduced from angular distributions measured at low energies ($E_{cm} <  1.7$ MeV). We show the new measurements  \cite{Kunz,Plag} together with the previous measurements \cite{Redder,Ouellet} that are not excluded here. Our analysis of angular distributions published in \cite{Kunz,Plag,Redder,Ouellet} confirms the published $S_{E2}$ cross section factors and error bars hence they are shown in Fig. 3 as published \cite{Kunz,Plag,Redder,Ouellet}. However, a few data points published with relative error bars of nearly 100\% (or larger) are not included in Fig. 3. The results obtained at Stuttgart \cite{Ham1,Ham2} are also not included in this sample of "world data" as discussed above.

The data shown in Fig. 3 aggregate into two distinct groups. On the one hand the R-matrix fit of Plag. {\em et al.} \cite{Plag} shown in Fig. 3 yields a reasonable fit ($\chi^2/N \ = $ 2.0, N=10) to the data measured by Plag {\em et al.} \cite{Plag} and Redder {\em et al.} \cite{Redder} but it yields a poor fit ($\chi^2/N \ = \ $ 6.0, N=5) to the data of both Kunz {\em et al.} \cite{Kunz} and Oulellet {\em et al.} \cite{Ouellet}. This fit extrapolates to $S_{E2}(300) \ = \ 60 \pm 12$ keVb. 

On the other hand the R-matrix fit curve of Plag {\em et al.} \cite{Plag}, when multiplied by 2.57, yields a good fit ($\chi^2/N \ = \ $ 0.75) to the data of Kunz {\em et al.} \cite{Kunz} and Oulellet {\em et al.} \cite{Ouellet}, but the so renormalized curve yields a poor fit ($\chi^2/N \ = \ $ 26.7) to the data of both Plag {\em et al.} \cite{Plag} and Redder {\em et al.} \cite{Redder}. This fit extrapolates to $S_{E2}(300) \ = \ 154 \pm 31$ keVb. The multiplicative factor of 2.57 may indeed reflect our lack of knowledge of, for example, the alpha-width (spectroscopic factor) of the bound $2^+$ state at 6.917 MeV in $^{16}$O.

\begin{figure}
\includegraphics[width=3.5in]{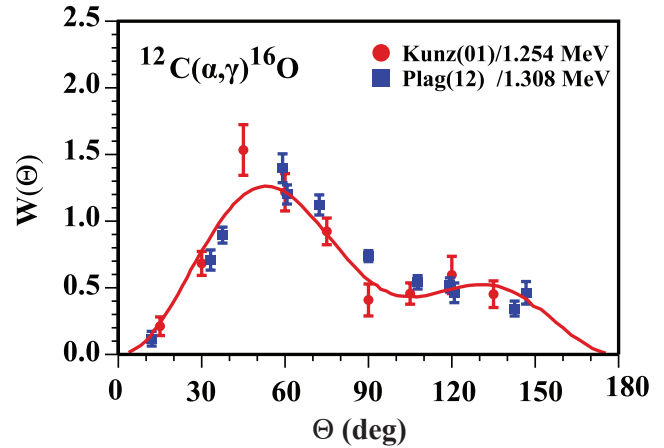}
\caption{\label{no3}(Color Online) The angular distributions measured by Plag {\em et al.} \cite{Plag} superimposed on the published data and fit curve of Kunz {\em et al.} \cite{Kunz}.}
\end{figure}

It is worth noting the subtle difference between the angular distribution published by Kunz  {\em et al.} \cite{Kunz} at $E_{cm} \ = \ 1.254$ MeV with $E2/E1$ = 1.28(32) as compared to the recent angular distribution published by Plag {\em et al.} \cite{Plag} at the nearby energy of $E_{cm} \ = \ 1.308$ MeV with $E2/E1$ = 0.44(20). The subtle differences in the data around $90^\circ$ lead to a factor of almost 3 in the value of the extracted $E2$ cross section. Such a large difference is clearly not expected due to the 54 keV difference in energy where these two angular distribution were measured and indicates a major systematic problem. Clearly, these recent angular distributions \cite{Kunz,Plag} are the most accurate data available today on the $^{12}$C$(\alpha,\gamma)$ reaction at low energy, but the two different  extrapolated values do not allow us to determine $S_{E2}(300)$ with the required accuracy of 10\% or better.

We conclude that current "world data" on $S_{E2}$ extracted from angular distributions measured at energies below 1.7 MeV cluster into two distinct groups leading to two different extrapolations of $S_{E2}(300) \ \approx 60$ or $\approx 154$ keVb. Neither one of these solutions can be favored or ruled out by the current "world data" of measured angular distributions. In order to resolve this ambiguity in the value of $S_{E2}(300)$ one needs to measure complete and very detailed gamma-ray angular distributions for the $^{12}$C$(\alpha,\gamma)$ reaction with high accuracy (with binning of 10$^\circ$ or less) at very low energies (below 1.5 MeV). As shown in Fig. 1 the data at large backward angles are most sensitive to the $E2/E1$ ratio, but such measurements with gamma-ray detectors are challenged by the finite size of the gamma-ray detector and the presence of the beam pipe.

The ambiguity in the value of the extrapolated $S_{E2}(300)$, reported in this paper for the first time, resembles the ambiguity in the value of the extrapolated $S_{E1}(300)$ value where even the data on the beta-decay of $^{16}$N shown in Fig. 18 (and Fig. 16) of \cite{TRIUMF} reveal two minima with identical $\chi^2_\beta$ values at $S_{E1}(300) \ \approx$ 10 keVb and $\approx$ 80 keVb. The small value of the extrapolated $S_{E1}(300) \approx 10$ keVb has been discussed by many authors \cite{Caltech,Redder,Ouellet2,Hale,Gial1} and cannot be resolved by the modern data as shown in Fig. 5 of \cite{Ham1}. In order to resolve this ambiguity in the value of $S_{E1}(300)$ the newly proposed experiments \cite{Gai,LUNA-MV} must measure complete gamma-ray angular distributions of the $^{12}$C$(\alpha,\gamma)$ reaction with high accuracy at low energies (below 1.0 MeV).

The Legendre-polynomial fit of the angular distribution data discussed above also includes an $E1-E2$ interference term with a ($\phi_{12}$) relative phase angle. This phase angle can be written as \cite{Barker}:
\begin{center}
$\phi_{12} \ = \ \delta_2 \ - \ \delta_1$ + arctan($\eta /2) \ \ \ \ \ \ (1)$
\end{center}
\    \\
where $\delta_1$ and  $\delta_2$ are the measured elastic phase shifts for $\ell$ = 1 and $\ell$ = 2 respectively, and $\eta$ is the Sommerfeld parameter. Since this relationship was first derived in (multi-level) R-matrix theory \cite{Barker} it is generally assumed to be a prediction of the R-matrix theory. But in fact the broader validity of equ. (1) was discussed in Ref. \cite{Brune} and it was previously shown to be a consequence \cite{Knutson} of the Watson theorem \cite{Watson} which itself is routed in unitarity. Hence we conclude that equ. (1) is required by unitarity.

The recently measured angular distributions were analyzed by either fixing the value of the $E1-E2$ mixing angle ($\phi_{12}$) at the value predicted by equ. (1) \cite{Kunz,Ham2} or by considering the phase angle ($\phi_{12}$) as a fit parameter \cite{Ham2}. The $E1-E2$ relative phases ($\phi_{12}$) extracted as fit parameters \cite{Ham2} are in strong disagreement with the prediction of equ. (1), as shown in Fig. 11 of \cite{Ham2}. Hence we conclude that the relative phase angles measured in the Stuttgart experiment \cite{Ham2} violate unitarity. Such strong deviations from equ. (1) are observed on resonance around $E_{cm}$ = 2.4 MeV where the cross sections are large. They indicate poorly understood systematic problems in the measured angular distributions \cite{Ham2}, as also concluded in \cite{Sayre}. Clearly this violation of unitarity must be resolved by future measurements of complete angular distribution measured in the vicinity of the $1^-$ resonance state of $^{16}$O ($E_{cm} \ \approx \ 2.4$ MeV).

To conclude we analyzed and reviewed new modern measurements of complete angular distributions of gamma-rays from the $^{12}$C$(\alpha,\gamma)$ reaction measured at very low energies approaching $E_{cm}  \approx  1.0$ MeV. While these measurements represent a major improvement of the "world data" and our knowledge of the low energy cross section of the $^{12}$C$(\alpha,\gamma)^{16}$O reaction, we demonstrate that thus far the measured $S_{E2}$ values bifurcate into two groups extrapolating to $S_{E2}(300) \ \approx$ 60 keVb or $\approx$154 keVb. This ambiguity in the extrapolated $S_{E2}(300)$ value resembles the ambiguity in the extrapolated $S_{E1}(300)$ value where the small $S_{E1}(300) \ \approx$ 10 keVb solution cannot be ruled out in favor of the large $\approx$ 80 keVb solution. These ambiguities in the extrapolated $S_{E2}(300)$ and $S_{E1}(300)$ values must be considered by practitioners in the field of stellar evolution theory and they must be resolved by the new generation experiments now in progress \cite{Gai} or in the planning stage \cite{LUNA-MV}. A violation of unitarity of the measured $E1-E2$ relative phases $(\phi_{12})$ must be resolved as well.
\vspace{-0.5cm}
\section{Acknowledgement}
\vspace{-0.5cm}
The author wish to acknowledge helpful discussion with Carl R. Brune and the help of Henry R. Weller in polishing this manuscript. This work was Supported by USDOE grant DE-FG02-94ER40870.

\end{document}